\documentclass[a4paper,12pt]{article}

\usepackage{amsmath}\usepackage{amssymb}
\usepackage{latexsym}\usepackage{cite}

\newcommand{\be}{\begin{equation}}
\newcommand{\ee}{\end{equation}}

\def\beq{\begin{equation}}
\def\eeq{\end{equation}}

\def\al{\alpha}
\def\bt{\beta}

\def\ga{\gamma}
\def\de{\delta}
\def\De{\Delta}

\def\Si{\Sigma}
\def\te{\theta}

\def\lam{\lambda}

\def\ep{\epsilon}

\def\sq{\sqrt}

\def\l{\left (}
\def\r{\right )}

\def\fr{\frac}
\def\la{\label}
\def\hs{\hspace}
\def\vs{\vspace}

\def\ran{\rangle}
\def\lan{\langle}
\def\ov{\overline}
\def\tl{\tilde}
\def\tm{\times}

\begin{document}

\begin{flushright}
BA-05-101\\
CERN-PH-TH/2005-164\\
\end{flushright}

\vs{0.5cm}

\begin{center}
{\Large\bf    

Supersymmetric $SO(10)$ And A Prediction For $\theta_{13}$}

\end{center}

\vspace{0.5cm}
\begin{center}
{\large 
{}~Qaisar Shafi$^{a}$\footnote{E-mail address: 
shafi@bartol.udel.edu}~~and~ 
{}~Zurab Tavartkiladze$^{b}$\footnote{E-mail address: 
Zurab.Tavartkiladze@cern.ch} 
}
\vspace{0.5cm}

$^a${\em Bartol Research Institute, Department of Physics and Astronomy

University of Delaware, Newark, DE 19716, USA \\

$^b$ Physics Department, Theory Division, CERN, CH-1211 Geneva 23, Switzerland
}

\end{center}
\vspace{0.6cm}

\begin{abstract}


We present a predictive scheme for fermion masses and mixings based on
supersymmetric $SO(10)$ which provides an understanding of the observed
hierarchies in the charged fermion  sector including CKM mixings. 
In the neutrino sector bi-large mixing can be realized, while the third 
leptonic mixing angle 
$\theta_{13}\simeq 0.01\hs{-0.3mm}-\hs{-0.3mm}0.2$. 

\end{abstract} 

\vs{0.7cm}


The $SO(10)$ gauge group has many attractive features for building Grand 
Unified Theories (GUT) \cite{Fritzsch:1974nn}. The standard model (SM) chiral 
fermions together with the  SM singlet right handed neutrinos (RHN) can be 
unified in the $16$ dimensional spinorial representations: 
$16\supset (q, u^c, d^c, l, e^c, \nu^c)$. The RHN states help generate non 
zero neutrino masses which can account for atmospheric \cite{Fukuda:2000np} 
and solar \cite{Fukuda:2001nj} neutrino oscillations. The masses of $\nu^c$ 
states arise from the breaking of lepton number and this ingredient proves 
decisive for generating the observed baryon asymmetry of the Universe through 
leptogenesis \cite{Fukugita:1986hr}. The supersymmetric (SUSY) version of 
$SO(10)$ strongly suggests unification of the SM gauge couplings at scale 
$M_G\simeq 2\cdot 10^{16}$~GeV.  

Many attempts have been made for building phenomenologically realistic models 
based on $SO(10)$ \cite{Lazarides:1980nt}-\cite{Aulakh:2003kg}.  
The problems associated with  $SO(10)$ model building arise both from 
the scalar  as well as the fermion sector. The scalar sector should be 
arranged to yield a suitable symmetry breaking pattern with preferably
(for MSSM unification) a one step symmetry reduction, 
$SO(10)\to SU(3)_C\tm SU(2)_L\tm U(1)_Y\equiv G_{321}$. At the same 
time, it is challenging to have natural doublet-triplet (DT) splitting in 
order to avoid  rapid nucleon decay, and also maintain successful unification 
of the three SM gauge couplings. For resolving the DT splitting problem
the mechanism in \cite{Dimopoulos:1981xm} (often called the missing VEV 
mechanism) can be invoked. Although its realization requires a complicated 
higgs sector supplemented by additional symmetries needed to avoid unwanted 
operators 
\cite{Babu:1993we}, \cite{Berezhiani:1996bv}, \cite{Kyae:2005vg},
it is a nice feature of  $SO(10)$ that DT splitting can be realized 
without fine tuning.

In this paper we do not consider the details of the scalar sector and 
concentrate mainly on the fermion sector. Therefore, our choice of the 
$SO(10)$ non-trivial scalar states will be as minimal as possible. 
{}For breaking SUSY $SO(10)$  down to $G_{321}$,
it is enough to introduce the scalar superfields 
$\Si (45)+C(16)+\bar C(\bar 16)$, where the brackets display the dimensions
of the $SO(10)$ representations. The rank reduction from five to four 
is achieved by the VEVs
$\lan C\ran =\lan \bar C\ran $ in  $\nu^c$ direction, while $\lan \Si \ran $
can be in a suitable direction preserving $G_{321}$.
To have a renormalizable Yukawa coupling for the top quark, we also introduce 
$H(10)$ state in which the light $h_u+h_d$ pair of MSSM higgs doublet
superfields reside. Note that, in principle, 
nothing prevents the $C$, $\bar C$ states from also contributing with some
weights to the physical doublet states \cite{Panagiotakopoulos:1985pt}. 
In fact, as we will see below, this 
can play an important role for realizing realistic pattern for fermion 
masses and mixings.

Turning to the fermion sector, let us note that the 
minimal version of SUSY $SO(10)$ (containing only three matter $16$-plets
and a single higgs $10$-plet)
gives the incorrect asymptotic relations 
$\hat{M}_U\propto \hat{M}_D=\hat{M}_E$,
$V_{CKM}={\bf 1}$ and is therefore excluded. To build a realistic model some 
extension is needed and numerous attempts have been considered 
(for an incomplete list see \cite{Lazarides:1980nt}-\cite{Aulakh:2003kg}).
In this paper we wish to build a simple and economical scenario
with realistic fermion pattern, and which also accounts for bi-large
atmospheric and solar neutrino mixings.
The task is not trivial and to achieve our goal we first formulate the 
principles which serve us as a guide for model building:

1) The field content must be as economical as possible in order to keep the
model simple and maintain perturbativity up to the cut off scale.

2) We wish to have the cut off scale at or near 
$M_{\rm Pl}\simeq 2.4\cdot 10^{18}$~GeV, so there is a large enough gap 
between $M_G$ and the cut off. Therefore there is a large interval where 
$SO(10)$ is unbroken, and with a fixed field content one can check 
perturbativity of the unified gauge coupling  $\al_{SO(10)}$ 
up to $M_{\rm Pl}$, taking into account all relevant 
threshold corrections.  

3) All operators allowed by the symmetries must be taken into account,
so that the conclusions are robust.

4) Last but not least, we would like to make the selection of the states and 
symmetries in such a way as to limit the allowed couplings and  
maximize the predictions. The third leptonic mixing angle $\te_{13}$
is a particularly important target in this regard.

In the fermion sector, in addition to the three  $16$-plets, one can 
also introduce fermionic $10$-plet states.
The latter contain vector-like states with the quantum numbers of the 
left handed doublet and down type quark. 
Because of this, the $10$-plets can play an important role 
\cite{Berezhiani:1996bv}, \cite{Shafi:1999au}, \cite{Maekawa:2001uk}, 
in the generation of charged 
fermion masses and mixings. In addition, one may need to introduce 
(scalar or fermionic) singlet superfields. They can play an important role 
in generating the desirable hierarchies and in explaining neutrino 
oscillations.
In order to protect the hierarchies and create predictive power, 
we introduce an
$R$-symmetry which will play a central role in our considerations.

The scalar and fermionic states (involved in the charged fermion 
sector) are given in table 1, which also lists the $R$-charges of the 
corresponding state.  $S$ and $\bar S$ denote scalar superfields which are 
singlets under $SO(10)$, while $10$ and $10'$ are matter supermultiplets. 
The role of each  state and $R$-charge assignment should 
become clear shortly. Under $R$, the superfield $\phi_i$ and the 
superpotential respectively transform as 
$\phi_i\to e^{{\rm i}\al_i}\phi_i$ and $W\to e^{{\rm i}(2\al_3+\al_H)}W$. 
%
%
%
\begin{table} \caption{$R$-charges of introduced  states.}
 
\label{t:1} $$\begin{array}{|c||c|c|c|c|c|}
 
\hline 

\vs{-0.3cm}

&  &  &  & &   \\ 

\vs{-0.3cm}

 & \Si (45)  & C(16) &\bar C(\bar 16)  & S & \bar S \\

&  &  &  & &   \\ 

 \hline
 
\vs{-0.2cm} 

 &  &  &  & &    \\ 

\vs{-0.2cm}

R & \al_{\Si } &
\fr{1}{2}\al_H\hs{-0.1cm}-\hs{-0.1cm}\fr{1}{4}\al_S
\hs{-0.1cm}+\hs{-0.1cm}2\al_{\Si }  &
-\fr{1}{2}\al_H\hs{-0.1cm}-\hs{-0.1cm}\fr{3}{4}\al_S\hs{-0.1cm}+\hs{-0.1cm}
2\al_{\Si }  &\al_S & \al_{\ov{S}}   \\

 &  &  &  & &    \\ 

\hline \hline

\vs{-0.3cm}

 &  &  &  & &    \\ 

\vs{-0.3cm} 

&H(10)& 16_1 & 16_2, 16_3   & 10 & 10'    \\ 

&  &  &  & &    \\ 

\hline

\vs{-0.2cm}

&  &  &  & &    \\

\vs{-0.2cm} 

R& \al_H  & \al_3\hs{-0.1cm}-\hs{-0.1cm}3\al_{\Si }\hs{-0.1cm}+\hs{-0.1cm}
\fr{1}{2}\al_S & \al_3\hs{-0.1cm}-\hs{-0.1cm}\al_{\Si },~\al_3  &
\al_3\hs{-0.1cm}+\hs{-0.1cm}\fr{1}{2}\al_H\hs{-0.1cm}-\hs{-0.1cm}
\fr{3}{4}\al_S\hs{-0.1cm}-\hs{-0.1cm}\al_{\Si } &
\al_3\hs{-0.1cm}+\hs{-0.1cm}\fr{1}{2}\al_H\hs{-0.1cm}+\hs{-0.1cm}
\fr{3}{4}\al_S\hs{-0.1cm}-\hs{-0.1cm}\al_{\Si }   \\

&  &  &  & &    \\

\hline

\end{array}$$
 
\end{table}
%
%
Table 1 lists all non trivial $SO(10)$  representations, so 
we can immediately check whether perturbativity 
works or not up to $M_{\rm Pl}$. 

Assuming for simplicity that below $M_G$ we just have the MSSM field 
content, the additional states enter into play above 
$M_G$, so that
$\al^{-1}(\mu )=\al^{-1}(M_G)+\fr{3}{2\pi }\ln \fr{\mu }{M_G}$ 
(for $\mu >M_G$).
Taking $\al (M_G)\simeq 1/24.4$ we get $\al (M_{\rm Pl})\simeq 1/26.7$.
Therefore, we remain in perfect perturbative regime which ensures that 
the prediction for gauge coupling unification can be trusted and also all 
possible GUT threshold corrections are calculable. 
This allows one to introduce, if needed, additional superheavy vector-like
states. For instance, up to $8$ pairs of $16+\ov{16}$ would be allowed .
To fix the allowed field content from the requirement of 
perturbativity up to  $M_{\rm Pl}$, let us point out that it is possible 
to introduce several 
$45$ and $54$ states with masses $\stackrel{>}{_\sim }M_G$. 
{}For example, the allowed numbers which preserve 
perturbativity are $(n_{45},~n_{54})=(5, 0), ~(4,1),~(1, 3)$. Several 
scalar $45$-plets together with $54$-plet can play an important role
in the solution of DT splitting problem 
\cite{Babu:1993we}, \cite{Berezhiani:1996bv}, \cite{Kyae:2005vg}.
Larger representations such $126$ and $210$  
(which can play crucial roles in symmetry breaking and fermion mass 
matrices \cite{Aulakh:2003kg}) are not favored from this 
viewpoint because the $SO(10)$ gauge coupling blows up not far above 
the scale $M_G$, thus not allowing a sufficiently high cut off scale. 
We will not pursue a
detailed study of these issues here but only mention that the selection of 
a  high  cut off scale $M^*\approx M_{\rm Pl}$ is also important for  
suppressing operators such as $\l \fr{\Si }{M^*}\r^n F_{\mu \nu }^2$,
which introduce unknown threshold corrections \cite{Shafi:1983gz}. 
With $M^*\approx M_{\rm Pl}$, these corrections can be safely ignored.

The hierarchies between charged fermion Yukawa couplings and the CKM 
mixing elements can be parameterized by the parameter $\lam \simeq 0.2$
as follows: 
$$ 
\lam_t\sim 1~,~~~~~\lam_u:\lam_c:\lam_t\sim \lam^8:\lam^4:1~.
$$
$$
\lam_b\sim \lam_{\tau }\sim \fr{m_b}{m_t}\tan \bt ~,~~~~
\lam_d:\lam_s:\lam_b\sim \lam^4:\lam^2:1~.
$$
$$
\lam_e:\lam_{\mu }:\lam_{\tau }\sim \lam^5:\lam^2:1~.
$$
\beq
V_{us}\approx \lam ~,~~~~V_{cb}\approx \lam^2 ~,~~~~V_{ub}=\lam^4-\lam^3 ~,
\la{obsmass}
\eeq  
where the MSSM parameter $\tan \bt =\fr{v_u}{v_d}$. 
Our aim is to gain a natural understanding of this hierarchical pattern.
As expansion parameters, in addition to $\lam \simeq $($0.2$),
we exploit the dimensionless parameter
$\fr{M_G}{M_{\rm Pl}}\sim \lam^2 \sim 10^{-2}$. Powers of this  naturally 
appear from the Planck scale suppressed non renormalizable operators.
Note that we take 
$\lan C\ran =\lan \bar C\ran \sim \lan \Si \ran \simeq M_G$.
In addition we assume that the scalar components of the superfields $S$,
$\bar S$ also develop VEVs close to $M_G$. Thus, 
\beq
\fr{\lan S \ran }{M_{\rm Pl}}\sim \fr{\lan \bar S \ran }{M_{\rm Pl}}
\sim \fr{\lan C \ran }{M_{\rm Pl}}=
\fr{\lan \bar C\ran }{M_{\rm Pl}}\sim 
\fr{\lan \Si \ran }{M_{\rm Pl}}
\equiv \ep \sim 10^{-2}~.
\la{defeps}
\eeq

With the $R$-charge assignments  in table 1, the couplings of $16$-plets with
$H(10)$ are
\begin{equation}
\begin{array}{ccc}
 & {\begin{array}{ccc}
\hs{-0.6cm}16_1\hspace{0.3cm} & \hspace{0.3cm}16_2 \hspace{0.3cm} 
& \hspace{0.4cm}16_3
\end{array}}\\ \vspace{1mm}
 
\begin{array}{c}
16_1\vs{0.2cm} \\ 16_2\vs{0.2cm} \\ 16_3  
 \end{array}\!\!\!\!\!\hs{-0.2cm} &{\left(\begin{array}{ccc}
\fr{\Si^2\bar CC }{M_{\rm Pl}^4} & 0  & 0
\\  
0 &\l \fr{\Si }{M_{\rm Pl}}\r^2 &
\fr{\Si }{M_{\rm Pl}}
 \\
0  & \fr{\Si }{M_{\rm Pl}} &
1
\end{array}\right)H}~,
\end{array}  \!\!  ~~~~~
\label{16H16}
\end{equation}
where in each entry a dimensionless couplings of order unity is 
understood.
Due to the appearance of operators of varying dimensions, there will be 
hierarchical structure between masses and mixings of different generations. 
The selection given in table 1 reduces the number of parameters and 
allows one to make several predictions.

{}From (\ref{16H16}) the asymptotic up type quark Yukawas are
\beq
\lam_t\sim 1~,~~~\lam_u:\lam_c:\lam_t\sim \ep^4:\ep^2:1~.
\la{upY}
\eeq
Note  that the contribution of $SO(10)$  violating VEV $\lan \Si \ran $
is strongly suppressed for the third generation $16_3$-plet, so that
at $M_G$ we have
\beq
\lam_t=\lam_b=\lam_{\tau }~.
\la{tbtau}
\eeq
On the other hand, the effect of $\lan \Si \ran $ in the couplings involving 
second and first generations is strong. This means that the rotations arising
from up and down left handed quark flavors are misaligned. This gives 
non trivial contribution to the CKM mixing matrix. From (\ref{16H16}) we see 
that $V_{cb}\sim \ep $ has the correct magnitude. However, $V_{us}$ and 
$V_{ub}$ vanish at this stage and  the contributions to the first and second 
generation masses of charged leptons and down quarks are negligible. 
Thus, additional contributions are necessary. We now come to the important 
role played by $10$, $10'$-plets and $C$, $\bar C$ 
states\footnote{One can check that an extension with only  $10$-plets or 
$C$, $\bar C$ states cannot yield a satisfactory picture. It seems necessary 
to invoke both $10$, $10'$ and $C$, $\bar C$ states.}.
The relevant couplings (allowed by $SO(10)\tm R$ symmetry)  are
\beq
\fr{\Si^2 }{M_{\rm Pl}} 10\cdot  10' +
\fr{S}{M_{\rm Pl}}10C 16_2+
\fr{CC}{M_{\rm Pl}}16_1\cdot 16_2+  
\fr{\bar C}{M_{\rm Pl}}10'H16_2~.
\la{10coupl}
\eeq 
These four couplings completely determine the masses of light generation 
of down quarks and charged leptons as well as CKM mixing angles $V_{us}$ and
$V_{ub}$. In order to study the effects of these couplings, it is useful to 
work in terms of $SU(5)$ multiplets. With $16_i=10_i+\bar 5_i+1_i$,
$10=5+\bar 5$, $10'=5'+\bar 5'$, and substituting appropriate VEVs, the 
couplings (\ref{16H16}), (\ref{10coupl}) yield the following mass matrix 
relevant for down quark and charged lepton masses:
\begin{equation}
\begin{array}{ccccc}
 & {\begin{array}{ccccc} 
\hs{0.2cm}\bar 5_1 & \hspace{0.6cm}
\bar 5_2  
& \hspace{0.6cm}\bar 5_3 & \hspace{0.5cm}\bar 5
 & \hspace{0.7cm}\bar 5'
\end{array}}\\ \vspace{1mm}
 
\begin{array}{c}
10_1 \\ 10_2 \\ 10_3  \\5' \\ 5
 \end{array}\!\!\!\!\!\hs{-0.2cm} &{\left(\begin{array}{ccccc}

 \ep^4\bar h_H~ & \ep_C\bar h_C ~ &
 0~ &0 ~ & 0
\\  
 \ep_C\bar h_C~ &\ep^2\bar h_H~ &\ep \bar h_H~ &
\ep \bar h_C~ &\ep  \bar h_H
 \\
 0~ &\ep \bar h_H~ & \bar h_H~ & 0~ & 0
 \\
 0~&0~ &0~ &\ep \lan \Si \ran ~ &0 \\

 0~ &\ep \lan C\ran ~ &0~ &0~ & \ep \lan \Si \ran

\end{array}\right)}~,
\end{array}  \!\!  ~~~~
\label{DLbig}
\end{equation} 
where $\bar h_H$ and $\bar h_C$ denote $SU(5)$ $\bar 5$-plets coming from 
$H$ and $C$ superfields respectively. With $\ep_C$($\sim 10^{-2}$) we denote 
$\lan C\ran /M_{\rm Pl}$ in order to indicate (where appropriate) that there 
is no $SU(5)$ breaking effect in the corresponding coupling. 
We see that there is strong mixing between
$\bar 5_2$ and $\bar 5'$ states and it involves the $SU(5)$ violating
VEV $\lan \Si \ran $. This will ensure that the wrong asymptotic relations
$m_s=m_{\mu }$, $m_e=m_d$ (common for minimal $SO(10)$ and $SU(5)$ GUTs)
are avoided. Here we also assume that the light down  higgs doublet
$h_d$ resides in $H$ and $C$ (in $\bar 5^H$ and $\bar 5^C$ in terms of 
$SU(5)$):
\beq
H\stackrel{\supset }{_\sim }h_d~,~~~~C\supset \lam h_d ~.
\la{downhw}
\eeq

Upon integration of $5$, $\bar 5$, $5'$ and $\bar 5'$ states, 
and keeping in mind the content of $SU(5)$ plets and composition of
their tensor products $10\supset q, e^c$, $\bar 5\supset d^c, l$, 
$10\tm \bar 5\tm \bar 5_{H, C}=qd^ch_d+e^clh_d+\cdots $,  
(\ref{DLbig}) yields the following $3\tm 3$ mass matrices for down quarks 
and charged leptons,
\begin{equation}
\begin{array}{ccc}
 & {\begin{array}{ccc}
\hs{-0.2cm}d_1^c (l_1)~~ & d_2^c (l_2)
& d_3^c (l_3)
\end{array}}\\ \vspace{1mm}
 
\begin{array}{c}
q_1 (e_1^c) \\ q_2 (e_2^c) \\ q_3 (e_3^c)  
 \end{array}\!\!\!\!\!\hs{-0.2cm} &{\left(\begin{array}{ccc}

 \ep^4~~ &~~ \ep_C \lam ~~  &~~0
\\  
 \ep_C \lam  ~~&~~\ep ~~ &~~\ep  
 \\
 0 ~~&~~\ep ~~  &~~ 1 
 
\end{array}\right)h_d}~.
\end{array}  \!\!  ~~~~~
\label{DL}
\end{equation}
{}From (\ref{10coupl}) we see that $16_3$ does not couple with $10$, $10'$ 
states and therefore the relations in (\ref{tbtau}) are not affected. 
However, we have now generated  masses for the second generation
states with the required  hierarchies
\beq
\fr{m_s}{m_b}\sim \fr{m_{\mu }}{m_{\tau }}\sim \ep ~,~~~~~~~~~
(m_s\neq m_{\mu})~.
\la{smu}
\eeq
The CKM mixing angles are
\beq
V_{us}\sim \lam ~,~~~~~V_{ub}\sim V_{cb}\lam ~,~~~~~
{\rm with}~~~V_{cb}\sim \ep ~.
\la{cabibo}
\eeq

The $d$-quark and the electron acquire masses through mixings with their 
nearest neighbors respectively (see (\ref{DL})). This yields the following  
relations
\beq
m_d=|V_{us}|^2m_s~,~~~~~m_e=|V_{e\mu }|^2m_{\mu }~.
\la{derel}
\eeq
There is no $SU(5)$ symmetry violating effects in the 
$(1,2)$ and $(2,1)$ entries of (\ref{DL}). 
This leads to another predictive relation at the high scale,
\beq
\fr{m_e}{m_d}=\fr{m_s}{m_{\mu }}~.
\la{reldesmu}
\eeq
With $\fr{m_{\mu }}{m_s}\approx 3$ near $M_G$ \cite{Georgi:1979df}, 
we have the desirable asymptotic relation $\fr{m_e}{m_d}\approx 0.3$. 
Thus, with a relatively simple $SO(10)$ model we have obtained 
a realistic pattern of charged fermion masses and their mixings.

Finally, 
the leptonic mixing angles acquire small contributions from the charged lepton 
sector, to wit, 
$\te_{e\mu }\simeq \sq{\fr{m_e}{m_{\mu }}}, \te_{\mu \tau }\sim \ep $. 
Therefore,  the neutral lepton sector (involving right handed neutrinos) 
must play an important role in order to obtain a satisfactory explanation of 
the solar and atmospheric neutrino anomalies. As we will see now the 
$10$, $10'$ states introduced above are also important for recovering 
bi-large neutrino mixings.

\vs{0.3cm}

The Dirac mass matrix for $\nu ,\nu^c$  differs from the case when there are
only fermionic $16_i$-plets. The light lepton 
doublets by appropriate weights reside also in the $10$-plets. This is indeed 
crucial point for natural bi-large neutrino mixings in our model. 
The couplings 
in eqs. (\ref{16H16}), (\ref{10coupl}) yield the following matrix 
\begin{equation}
\begin{array}{ccccc}
 & {\begin{array}{ccccc}
\hs{-0.3cm}\nu^c_1\hspace{0.2cm} & \hspace{0.3cm}
\nu^c_2 \hspace{0.2cm} 
& \hspace{0.3cm}\nu^c_3 \hs{0.1cm} & \hspace{0.4cm}\bar l_{5'}
 & \hspace{0.5cm}\bar l_{5}
\end{array}}\\ \vspace{1mm}
 
\begin{array}{c}
l_1 \\ l_2 \\ l_3  \\l_{\bar 5} \\ l_{\bar 5'}
 \end{array}\!\!\!\!\!\hs{-0.2cm} &{\left(\begin{array}{ccccc}

 \ep^4h_H & 0  &0    &0&0   
\\  
 0 &\ep^2h_H &\ep h_H &0 &\ep \lan C\ran 
 \\
 0 &\ep h_H & h_H &0  &0
 \\
 0&0 &0 &\ep \lan \Si \ran  &0 \\

  0 &\ep h_H &0 &0 &\ep  \lan \Si \ran

\end{array}\right)}~,
\end{array}  \!\!  ~~~~~
\label{ML}
\end{equation}
where $h_H$  denotes the up-type higgs doublet coming from $H$. Integration 
of $l_5$, $l_{5'}$ states lead to the $3\tm 3$ mass matrix 
\begin{equation}  
\begin{array}{ccc}
 & {\begin{array}{ccc}
\hs{-0.6cm}\nu^c_1 & 
~\nu^c_2  &~ \nu^c_3 
\end{array}}\\ \vspace{1mm}
m_D=  
\begin{array}{c}
l_1 \\ l_2 \\ l_3  
 \end{array}\!\!\!\!\!\hs{-0.2cm} &{\left(\begin{array}{ccc}

 \ep^4 & ~0  &~0
\\  
 0 &~\ep &~\ep 
 \\
 0 &~\ep  &~ 1
 
\end{array}\right)h_H}~.
\end{array}  \!\!  ~~~~~
\label{mD}
\end{equation}
We see that (\ref{mD})  differs in structure from the up type quark mass 
matrix. In particular, the $(2,2)$ and $(3,2)$ elements in (\ref{mD}) 
are of the same order (in $\ep $). This offers a natural way for large 
$\nu_{\mu }-\nu_{\tau }$ mixing. However, we should also ensure the  
generation of the required value of 
$\De m_{\rm sol}^2/\De m_{\rm atm}^2\sim 1/5$,  i.e. at  leading 
order only one eigenstate should have mass close to 
$\sq{\De m_{\rm atm}^2}$\footnote{This requirement must be fulfilled because 
the neutrino masses turn out to be hierarchical.}.
This is possible to realize if effectively in the see-saw mechanism only one 
RHN exchange dominates (the single RHN dominance mechanism)
\cite{Suematsu:1996mk}, \cite{Shafi:1998dv}. For this to be realized the 
other two RHN states should decouple \cite{Shafi:1998dv}
without significant contribution to the 
light neutrino mass matrix. 

To achieve this, we introduce 
three $SO(10)$ singlet states $N_i$ ($i=1,2,3$) with $R$-charges
\beq
\al_{N_2}=\al_3+\fr{3}{2}\al_H-\al_{\Si }-\fr{1}{4}\al_S~,~~~
\al_{N_1}=\al_{N_2}+2\al_{\Si }+\fr{1}{2}\al_S~,~~~
\al_{N_3}=\al_{N_2}-\al_{\Si }+\al_S~.
\la{RNi}
\eeq 
Note that for the time being not all phases in table 1 and (\ref{RNi}) 
are fixed. Taking
\beq
\al_H=\al_{\Si }+\fr{1}{4}\al_{S}~,
\la{sel1}
\eeq
the couplings involving $N$-states are 
\begin{equation}
\begin{array}{ccc}
 & {\begin{array}{ccc}
\hs{-1cm}N_1\hspace{0.8mm} & N_2
 &\hs{0.8mm} N_3 
\end{array}}\\ \vspace{1mm}
 
\begin{array}{c}
16_1\vs{0.1cm} \\ 16_2\vs{0.1cm} \\ 16_3  
 \end{array}\!\!\!\!\!\hs{-0.2cm} &{\left(\begin{array}{ccc}

 1 &0  & 0
\\  
 0&\fr{S }{M_{\rm Pl}} &\fr{\Si }{M_{\rm Pl}}
 \\
 0&0& 1
 
\end{array}\right)\bar C}~,~~~
\end{array}  
\begin{array}{ccc}
 & {\begin{array}{ccc}
\hs{-0.7cm}N_1 &\hs{-0.1cm} N_2 & \hs{-0.1cm}N_3
\end{array}}\\ \vspace{1mm}
 
\begin{array}{c}
N_1 \\ N_2 \\ N_3  
 \end{array}\!\!\!\!\!\hs{-0.2cm} &{\left(\begin{array}{ccc}

 0 &~0 &~ 0
\\  
 0&~1 &~0
 \\
 0&~0&~ 0
 
\end{array}\right)M_{\rm Pl}}~.
\end{array}  \!\!  
\label{16N}
\end{equation} 
{}From eqs. (\ref{mD}), (\ref{16N}) the $9\tm 9$ mass matrix for neutral 
fermions is given by
\begin{equation}
\begin{array}{ccc}
 & {\begin{array}{ccc}
\hs{-0.15cm}\nu \hspace{0.2cm} & \hspace{0.15cm}
\nu^c \hspace{0.2cm} 
& \hspace{0.15cm}N 
\end{array}}\\ \vspace{1mm}
 
\begin{array}{c}
\nu \\ \nu^c \\ N
 \end{array}\!\!\!\!\!\hs{-0.2cm} &{\left(\begin{array}{ccccc}

 0 & m_D &\De 
\\
 m_D^T&0 &M
 \\
 \De^T&M^T & M_N
\end{array}\right)}~,
\end{array}  \!\!  ~~~~~
\label{nubig}
\end{equation}
where $M$ and $\De $ emerge from the first matrix coupling of (\ref{16N}),
while $M_N$ accounts for $NN$-type coupling. The $\De $ contribution is
non zero if the up type higgs doublet $h_{\bar C}$ from $\bar C$ has 
non zero VEV.
Integration of the heavy $\nu^c$, $N$ states induces contributions to the 
light neutrino mass matrix given by the double see-saw formula
\beq
m_{\nu }^{(0)}=m_D\fr{1}{M^T}M_N\fr{1}{M}m_D^T-\De\fr{1}{M}m_D^T-
m_D\fr{1}{M^T}\De^T~.
\la{doubleSS}
\eeq

To analyze this mass matrix it is convenient to work in a basis in which the 
first coupling matrix of (\ref{16N}) is diagonal. With this change of basis 
the structures of  $m_D$ and $M_N$ will be unchanged and we can 
parameterize the relevant matrices as
$$
\begin{array}{ccc}
 & {\begin{array}{ccc}
\hs{-0.1cm} & 
  & 
\end{array}}\\ \vspace{1mm}
m_D=  
\begin{array}{c}
 \\  \\  
 \end{array}\!\!\!\!\!\hs{-0.2cm} &{\left(\begin{array}{ccc}

 \ep^4 & 0  &0
\\  
 0 &\bt \ep &\al \ep 
 \\
 0 &\al \ep  & 1
 
\end{array}\right)h_H}~,
\end{array}  \!\!  
$$
\beq
\fr{1}{M}={\rm Diag}\l \ep ,~n_1,~\ep \r \cdot \fr{1}{M_{\rm Pl}\ep^2} ~,~~~
M_N={\rm Diag}\l 0,~n_2,~0\r M_{\rm Pl}~,
\la{parmat}
\eeq
where the dimensionless couplings
$\al, \bt ,n_{1,2}$ of order unity have been identified in order to 
demonstrate that our goal (for which $N$-states have been introduced) is 
achieved.

Employing (\ref{doubleSS}), the dominant contribution to the $3\tm 3$ 
neutrino mass matrix is given by
\beq
\begin{array}{ccc}
 & {\begin{array}{ccc}
\hs{-1cm} &  &\,\,~~~
\end{array}}\\ \vspace{2mm}
\!\!\!\!\! &~ {m_{\nu }^{(0)}
= \left(\hs{-0.2cm}\begin{array}{ccc}
~0&~0 &~0 
\\
~0 & ~\bt^2 &~ \al \bt
\\
~0 &~ \al \bt &~\al^2
\end{array}\hs{-0.15cm}\right)\fr{m}{|\al |^2+|\bt |^2}~,~~~~~ }
{\rm with}~~~m=(|\al |^2+|\bt |^2)n_1^2n_2\fr{(h_H^0)^2}{M_{\rm Pl}\ep^2}~.
\end{array}  
\la{m0}
\eeq  
We see that only one eigenstate  of (\ref{m0}) is massive with 
$m_{\nu_3}=m\sim \fr{\lan h_H^0\ran^2}{M_{\rm Pl}\ep^2}=0.01-0.1$~eV -
the scale relevant for  atmospheric neutrino oscillations. 
Moreover,
the 2-3 mixing angle $\te_{23}$ is naturally large for 
$\al \hs{-0.1cm}\sim \hs{-0.1cm}\bt \hs{-0.1cm}\sim \hs{-0.1cm}1$.  
The form in (\ref{m0}) emerges from the first term in 
(\ref{doubleSS}). The second and third terms of (\ref{doubleSS}) 
induce (negligible) corrections irrelevant for neutrino oscillations. 

{}For explaining 
the solar neutrino anomaly, we introduce an additional singlet 
state ${\cal N}$ with the $R$-charge 
$\al_{\cal N}=\al_3+\fr{1}{2}\al_H+2\al_{\Si }-\fr{3}{2}\al_S$.
Note that we still have some freedom in selection of phases (determining 
the $R$-charges), and we use this freedom to take 
$\al_{\Si }=\fr{29}{48}\al_S ,\al_{\ov{S} }=\fr{1}{12}\al_S $. 
With  this and the prescriptions in table 1,
and noting (\ref{sel1}), the relevant superpotential couplings are
\beq
\bar S{\cal N}10' \bar C\bar C+S^2{\cal N}16_1\bar C+\bar S^7{\cal N}^2~,
\la{solNcoupl}
\eeq
where we have omitted appropriate powers of $M_{\rm Pl}$ which can be 
restored when needed.
Note that the $10'$-plet contains the $l_2$ ($\nu_2$) state with weight
$\sim 1$. So, the coupling of $10'$ in (\ref{solNcoupl}) is of great 
importance.  
With this setting  the coupling 
$\bar S\Si {\cal N}16_3\bar C$ is also allowed.Since this coupling is 
relevant  we will take it into account for generality. 
Substituting appropriate VEVs and integrating the ${\cal N}$ state we get 
a sub dominant contribution to the light neutrino mass matrix
\begin{equation}
\begin{array}{cc}
& {\begin{array}{ccc}
\hs{-1.2cm}~ &~  &~
\end{array}}\\ \vspace{2mm}
\begin{array}{c}
  \\ \\ 

\end{array} &{m_{\nu }^{(1)}=\left(\begin{array}{ccc}
\bar \al^2 &~~~\bar \al \bar \bt &~~~\bar \al \bar \ga 
\\
\bar \al \bar \bt &~~~\bar \bt^{2}&~~~\bar \bt \bar \ga 
\\
\bar \al \bar \ga  &~~~\bar \bt \bar \ga &~~~\bar \ga^2 
\end{array}\hs{-0.2cm}\right)m' ~,
}
\end{array}
\label{m1}
\end{equation}    
where the dimensionless couplings $\bar \al , \bar \bt , \bar \ga  \sim 1$ 
and
\beq
m'\sim \fr{\lan h_{\bar C}\ran^2}{M_{\rm Pl}\ep^3}~.
\la{mpr}
\eeq
We already see that the  mixing angle $\te_{12}$ is naturally large for 
$\bar \al\sim \bar \bt \sim 1$. As far as the mass scale $m'$ is concerned, 
assuming the light physical up type higgs doublet is contained in $H$ 
and $\bar C$ with weights
\beq
H\supset h_u~,~~~~\bar C\supset \sq{\fr{\ep }{5}}h_u~,
\la{upweights}
\eeq
we have 
$m'=2\hs{-0.3mm}\cdot \hs{-0.3mm}10^{-3}-5\hs{-0.3mm}\cdot \hs{-0.3mm}10^{-2}$~
eV, which accounts for solar neutrino oscillations.

Next we  study in detail the full neutrino mass matrix. Although in the 
neutrino sector there were many unknown parameters, the effective 
$3\tm 3$ neutrino mass matrix has a relatively simple structure.
Because of this, the model enables one to calculate the third mixing angle 
$\te_{13}$ in terms of other measured oscillation parameters. Once more we 
emphasize that this is possible thanks to the fermionic $10$-plets.

The  neutrino mass matrix can be written as
\beq
m_{\nu }=m_{\nu }^{(0)}+m_{\nu }^{(1)}~,
\la{mnu}
\eeq
where $m_{\nu }^{(0)}$ denotes the dominant part responsible for the 
atmospheric anomaly, while the sub-dominant entry $m_{\nu }^{(1)}$ ensures 
large angle solar neutrino oscillations. They are given in (\ref{m0}) and 
(\ref{m1}), with  $m'/m\ll 1$. The leading part can be diagonalized by the 
transformation  
$U_{23}^Tm_{\nu }^{(0)}U_{23}=\ov{m}_{\nu }^{(0)}={\rm Diag}\l 0,~0,~m\r $,
with
\beq
\begin{array}{ccc}
 & {\begin{array}{ccc}
 & & 
\end{array}}\\ \vspace{2mm}
U_{23}=
\begin{array}{c}
\end{array}\!\!\!\!\!\! &P{\left(\begin{array}{ccc}
\hs{-0.1cm}1~,&
0 ~, &0
\\
\hs{-0.2cm}0~, &
\hs{-0.1cm}c_{23}~,& s_{23}
\\
\hs{-0.1cm} 0~,&-s_{23} ~,&
c_{23}
\end{array} \right)\! }~,
\end{array}  
\la{U23}
\eeq
where $c_{23}\equiv \cos \te_{23}$, $s_{23}\equiv \sin \te_{23}$, and
\beq
\tan \te_{23}=\fr{|\bt |}{|\al |}~,~~~~
P={\rm Diag}\l 1,~e^{{\rm i}\cdot \chi },~1 \r ~,~~~
\chi =-{\rm Arg}(\al  \bt )~.
\label{PU23}
\end{equation} 
By this rotation, the sub-leading part transforms into 
$$
\begin{array}{ccc}
 & {\begin{array}{ccc}
~ & &\,\,~~~
\end{array}}\\ \vspace{2mm}
\begin{array}{c}
 \\  \\ 
\end{array}\!\!\!\!\! &\ov{m}_{\nu }^{(1)}=U_{23}^Tm_{\nu }^{(1)}U_{23}
= {\left(\begin{array}{ccc}
\bar \al^2& \bar \al \tl{\bt }  &
\bar \al \tl{\ga }
\\
\bar \al \tl{\bt} &  
\tl{\bt }^2 &
\tl{\bt }\tl{\ga }
\\
 \bar \al \tl{\ga } & 
\tl{\bt }\tl{\ga }
  & \tl{\ga }^2
\end{array}\right)\cdot m' } ~,
\end{array}  
\hs{-1cm}
$$
\beq
{\rm with }~~~\tl{\bt }=(\bar \bt c_{23}e^{{\rm i}\chi }-\bar \ga s_{23})~,~~
\tl{\ga }=(\bar \bt s_{23}e^{{\rm i}\chi }+\bar \ga c_{23})~.
\label{bmnu1}
\end{equation}    
{}From (\ref{bmnu1}) we see that for $\bar \al \sim \bar \bt $ the angle
$\te_{12}$ is naturally large and therefore  bi-large neutrino mixing 
is realized. On the other hand, the third mixing angle 
$\te_{13}\sim \fr{m'}{m}\sim \sq{\fr{\De m_{\rm sol}^2}{\De m_{\rm atm}^2}}$
is properly suppressed\footnote{A similar relation was obtained in a 
democratic scenario \cite{Shafi:2004jy} with discrete symmetries. 
We note, however, that so far it has not been generated within a 
GUT scenario supplemented by symmetry arguments.}.

The model allows one to make one more prediction if somehow the coupling 
${\cal N}16_3\bar C$ can be forbidden. This is realized with a suitable 
modification of $R$-charges. For example, if the last coupling in 
(\ref{10coupl}) is generated through the integration of some additional 
states. Introduce the vector-like fermionic pair $F(16)+\bar F(\bar 16)$
and, instead of the last term in (\ref{10coupl}), consider the couplings
\beq
16_2FH+\fr{S}{M_{\rm Pl}}10'\bar F\bar C+\bar S\bar FF~.
\la{Fcoupl}
\eeq
One can verify that with substitution of appropriate VEVs and
upon integration of $F, \bar F$ states, the operator
$\fr{\lan S\ran \bar C}{\lan \bar S\ran M_{\rm Pl}}10'H16_2\sim
\fr{\bar C}{M_{\rm Pl}}10'H16_2$ is 
generated, so that all results obtained in the charged fermion sector are 
robust. However, with the couplings in (\ref{Fcoupl}) the corresponding 
$R$-charges are modified as follows:
$$
\al_{C}\to \al_C-\fr{1}{4}(\al_S-\al_{\ov{S}})~,~~
\al_{\ov{C}}\to \al_{\ov{C}}-\fr{3}{4}(\al_S-\al_{\ov{S}})~,~~
\al_{16_1}\to \al_{16_1}+\fr{1}{2}(\al_S-\al_{\ov{S}})~,
$$
$$
\al_{10}\to \al_{10}+\fr{1}{4}(\al_S-\al_{\ov{S}})~,~~
\al_{10'}\to \al_{10'}-\fr{1}{4}(\al_S-\al_{\ov{S}})~,~~
\al_{N_i}\to \al_{N_i}+\fr{3}{4}(\al_S-\al_{\ov{S}})~,
$$
\beq
\al_{\cal N}\to \al_{\cal N}+\fr{1}{4}(\al_S-\al_{\ov{S}})~.
\la{modR} 
\eeq
The $R$-charges of states which are not given here are unchanged. 
With this changes and with the following relations between phases:
\beq
\al_H=\al_{\Si }-\fr{1}{2}\al_S+\fr{3}{4}\al_{\ov{S}}~,~~
\al_{\Si }=\fr{79}{64}\al_S~,~~~
\al_{\ov{S}}=-\fr{3}{8}\al_S~,
\la{relpas1}
\eeq
all of the couplings presented above, with exception of ${\cal N}16_3\bar C$, 
survive. With 
$\bar \ga =0$,  (\ref{bmnu1}) reduces to 
\begin{equation}
\begin{array}{ccc}
 & {\begin{array}{ccc}
~ & &\,\,~~~
\end{array}}\\ \vspace{2mm}
\begin{array}{c}
 \\  \\ 
\end{array}\!\!\!\!\! &\ov{m}_{\nu }^{(1)}=U_{23}^Tm_{\nu }^{(1)}U_{23}
= \tl{P}\cdot {\left(\begin{array}{ccc}
\bar \al^2 & c_{23}\bar \al \bar \bt  & s_{23}\bar \al \bar \bt
\\
c_{23}\bar \al \bar \bt & c_{23}^2\bar \bt^{2} & c_{23}s_{23}\bar \bt^{2}
\\
s_{23}\bar \al \bar \bt & c_{23}s_{23}\bar \bt^{2} & s_{23}^2\bar \bt^{2}
\end{array}\right)\cdot \tl{P}m' } ~,
\end{array}  
\hs{-1cm}
\label{mnu0}
\end{equation}    
where 
$\tl{P}={\rm Diag}(1, ~e^{-{\rm i}\cdot \chi },~e^{-{\rm i}\cdot \chi })$.

The matrix 
$\ov{m}_{\nu }=\ov{m}_{\nu }^{(0)}+\ov{m}_{\nu }^{(1)}$ 
is diagonalized by the transformation 
$U_{12}^TU_{13}^T\ov{m}_{\nu }U_{13}U_{12}=m_{\nu }^{\rm diag}$, where
$$
\begin{array}{ccc}
 & {\begin{array}{ccc}
 & & 
\end{array}}\\ \vspace{2mm}
U_{12}\simeq P'\cdot
\begin{array}{c}
\end{array}\!\!\!\!\!\! &{\left(\begin{array}{ccc}
\hs{-0.1cm}c_{12}~,&
s_{12} ~, &0
\\
\hs{-0.2cm}-s_{12}~, &
\hs{-0.1cm}c_{12}~,& 0
\\
\hs{-0.1cm} 0~,&0 ~,&
1
\end{array} \right)\! }~,
\end{array}  
~~~
\begin{array}{ccc}
 & {\begin{array}{ccc}
 & & 
\end{array}}\\ \vspace{2mm}
U_{13}\simeq 
\begin{array}{c}
\end{array}\!\!\!\!\!\! &{\left(\begin{array}{ccc}
\hs{-0.1cm}c_{13}e^{{\rm i}\cdot \de }~,&
0 ~, &s_{13}e^{{\rm i}\cdot \de }
\\
\hs{-0.2cm}0~, &
\hs{-0.1cm}1~,& 0
\\
\hs{-0.1cm}-s_{13}~,&0 ~,&
c_{13}
\end{array} \right)\! }~,
\end{array}  
$$
\beq
P'={\rm Diag}\l e^{{\rm i}\cdot \chi'},~e^{-{\rm i}\cdot \chi },~
e^{-{\rm i}\cdot \chi } \r ~,~~~\chi'=-{\rm Arg}(\bar \al \bar \bt )~,
~~~\de =-{\rm Arg}\l \fr{m'\bar \al \bar \bt }{m}\r ~,
\label{U12}
\end{equation} 

\beq
\tan \te_{12}=\fr{|\bar \al |}{c_{23}|\bar \bt |}~,~~~~~
\tan \te_{13}^{\nu }=\left |\fr{m'}{m}\right |s_{23}|\bar \al \bar \bt |~,
\la{tans}
\eeq

\beq
m_{\nu }^{\rm diag}={\rm Diag}(0, m_2, m)~,~~~~~
m_2=m'(|\bar \al |^2+c_{23}^2|\bar \bt |^2)~.
\la{mnudiag}
\eeq 
{}For the $\te_{13}^{\nu }$ mixing angle we have introduced the superscript
'$\nu $' in order to indicate that this contribution comes from the 
neutrino sector. As we will see, the whole leptonic mixing angle
$\te_{13}\equiv |U_{e3}^l|$ receives  sizable contribution also from 
the charged lepton sector. 
Using (\ref{tans}), (\ref{mnudiag}) and the relation
$\left |\fr{m_2}{m}\right |\simeq 
\sq{\fr{\De m_{\rm sol}^2}{\De m_{\rm atm}^2}}$, we find
\beq
\tan \te_{13}^{\nu }\simeq \sq{\fr{\De m_{\rm sol}^2}{\De m_{\rm atm}^2}}
\fr{\tan \te_{12}\tan \te_{23}}{1+\tan^2 \te_{12}}~.
\la{13pred}
\eeq
Using the current data \cite{Fogli:2005gs}, we find 
\beq
\te_{13}^{\nu }\simeq 0.05-0.14~.
\la{13range}
\eeq
{}For the central values 
$$
\De m_{\rm sol}^2=7.9\cdot 10^{-5}{\rm eV}^2~,~~~~
\De m_{\rm atm}^2=2.4\cdot 10^{-3}{\rm eV}^2~,
$$
\beq
\sin^2\te_{12}=0.314~,~~~~\sin^2\te_{23}=0.44~,
\la{data}
\eeq
(\ref{13pred}) yields
$(\te_{13}^{\nu })^{\rm cent}\simeq 0.075$. 
Now we can derive the $(1,3)$ element of the leptonic mixing matrix $U^l$.
A sizable contribution from the charged lepton sector comes only for
$\te_{13}\equiv |U_{e3}^l|$. Namely, the charged lepton $(1,2)$ 
mixing angle $\te_{e\mu }\simeq \sq{\fr{m_e}{m_{\mu }}}$ is relevant. 
Taking this into account, we obtain
\beq
\te_{13}\equiv |U_{e3}^l|\simeq \left | \tan \te_{13}^{\nu}-
\sq{\fr{m_e}{m_{\mu }}}s_{23}e^{{\rm i}(\de -\chi )}\right | 
\cos \te_{13}^{\nu }~,
\la{Ue3}
\eeq
where angles and phases are defined in (\ref{PU23}), 
(\ref{U12})-(\ref{13pred}). Since the phase $\de -\chi $ is unknown, we 
can calculate a range for $|U_{e3}^l|$. Namely, from (\ref{Ue3}),
$$
{\rm for}~~~\te_{13}^{\nu }=0.05~,~~~~\te_{13}=0.01-0.09~,
$$
\beq
{\rm for}~~~\te_{13}^{\nu }=0.14~,~~~~\te_{13}=0.08-0.2~.
\la{e3ranges}
\eeq
Thus, finally we get\footnote{Numerically similar values have been recently 
predicted in \cite{Ferrandis:2004mq}.} 
\beq
\te_{13}=0.01-0.2~.
\la{Ue3pred}
\eeq
The upper range in  (\ref{Ue3pred}) is consistent with the
current experimental bound \cite{Apollonio:2002gd}, while future
experiments \cite{Itow:2001ee} should be able to probe values
close to $0.01$.

\vs{0.5cm}

In summary we have proposed a relatively simple extension 
of minimal $SO(10)$ by introducing two 'matter' $10$-plets 
(and possibly one pair of vector-like matter $16+\ov{16}$
allowing us to predict the value of $\te_{13}$) plus some
$SO(10)$ singlets. Augmented by a $R$-symmetry these states play 
an essential role in understanding the hierarchies in the charged fermion  
sector and in realizing bi-large mixing in the neutrino sector. 
The third leptonic mixing angle turns out to be naturally suppressed , 
in the range of $0.01-0.2$. The $R$-symmetry also implies 'matter' 
parity, so that the LSP is stable. It also forbids Planck scale suppressed 
baryon number violating dimension five operators. 
Since $R$-symmetry has previously been shown to play an essential role
in the construction of realistic $SO(10)$ inflation models
\cite{Kyae:2002hu}, \cite{Kyae:2005vg}, it would be interesting to try to 
merge the two approaches and also include leptogenesis. This will be 
attempted elsewhere.

\bibliographystyle{unsrt}

\end{document}